# Automatic and quantitative measurement of spectrometer aberrations

Yueming Guo[1*], Andrew R. Lupini[1*]

1. Center for Nanophase Materials Sciences, Oak Ridge National Laboratory, Oak Ridge, TN 37830

*corresponding authors: dr.yueming.guo@gmail.com and arl1000@ornl.gov

*Notice: This manuscript has been authored by UT-Battelle, LLC, under Contract No. DE-AC0500OR22725 with the U.S. Department of Energy. The United States Government retains and the publisher, by accepting the article for publication, acknowledges that the United States Government retains a non-exclusive, paid-up, irrevocable, world-wide license to publish or reproduce the published form of this manuscript, or allow others to do so, for the United States Government purposes. The Department of Energy will provide public access to these results of federally sponsored research in accordance with the DOE Public Access Plan (http://energy.gov/downloads/doe-public-access-plan).*

**Abstract**

The performance of electron energy-loss spectrometers can often be limited by their electron-optical aberrations. Due to recent developments in high energy-resolution and momentum-resolved electron energy loss spectroscopy (EELS), there is renewed interest in optimizing the performance of such spectrometers. For example, the "ω-q" mode of momentum-resolved EELS, which uses a small convergence angle and requires aligning diffraction spots with a slot aperture, presents a challenge for realigning the spectrometer after adjusting the projection lenses. Automated and robust alignment can greatly benefit such a process. The first step towards this goal is automatic and quantitative measurement of spectrometer aberrations. Here we demonstrate the measurement of geometric aberrations and distortions in EELS within a monochromated scanning transmission electron microscope (STEM). To better understand the results, we present a wave mechanical simulation of the experiment. Using the measured aberration and distortion coefficients as inputs to the simulation, we find a good match between the simulation and experiment, verifying the approach used in the simulation, allowing us to assess the accuracy of the measurements. Understanding the errors and inaccuracies in the procedure can guide further progress in aberration measurement and correction for new spectrometer developments.

**Introduction**

Electron energy loss spectroscopy(EELS) is one of the foremost tools for atomic scale analysis of new materials, able to provide atomic column resolution (Batson 1993; Browing et al. 1993; Muller et al. 1993; Allen et al. 2003; Bosman et al. 2007; Ramasse et al. 2009; Botton et al. 2010) and even single-atom sensitivity (Varela et al 2004). To tackle currently relevant problems, such as topological materials, the experimental need is to probe the electronic structure with high-fidelity in an appropriate combination of energy, real space (position) and reciprocal space (momentum) resolved conditions.

Developments in monochromators (Terauchi et al. 1999; Tiemeijer 1999; Mukai et al. 2006; Krivanek et al. 2009) have produced a new generation of aberration-corrected instruments that enable high energy resolution with atomic sized probes. Recent advances in monochromator and spectrometer designs in a scanning transmission electron microscope (STEM) have improved the energy resolution to below 10 meV (Krivanek et al. 2014) with the present state-of-the-art at 3 meV at 20 kV (Dellby et al. 2020). This progress has enabled a variety of new applications. Notable examples include mapping of phonon density of states with high spatial resolution (Dwyer et al. 2016; Lagos et al. 2017, Venkatraman et al. 2019, Yan et al. 2021), mapping of phonon dispersion at nanoscale (Hage et al. 2018; Qi et al. 2022; Gadre et al. 2022), determination of local temperature (Idrobo et al. 2018), bonding analysis and detection of hydrogen (Miyata et al. 2013, Cohen et al. 2015, Haiber *et al.*2018), and probing beam-sensitive samples with vibrational fingerprints in an aloof mode (Hachtel *et al.* 2019).

The success of ultrahigh energy resolution in STEM has relied on improved instrumental stabilities (Krivanek et al. 2009; Krivanek et al. 2013) and aberration correction in electron optics, with a detailed discussion given in (Krivanek *et al.* 2019). It appears that further improvements in energy resolution will depend on both of these factors as well as mitigation of the Johnson noise (Uhlemann *et al.* 2013) and implementation of new detectors (Tate et al. 2016; Mir et al. 2017).

Although some progress for automatic aberration correction in monochromated EELS has been made by Nion Co. as evidenced by a python program within their software package, Nionswift, (Meyer *et al.* 2019) and briefly mentioned in (Krivanek *et al.* 2019), a more thorough effort is still lacking. In parallel, progress in aberration measurement and correction for EELS in general (regardless of monochromation) has also been made by CEOS (Kahl et al. 2019) and Gatan (Twesten *et al.* 2011). Early efforts in EELS aberration measurement and correction should also be acknowledged (Plies & Rose 1971; Rose & Plies 1973; Egerton 1980; Isaacson & Scheinfein 1983; Scheinfein & Isaacson 1984; Krivanek et al. 1987; Krivanek et al. 1991; Yang & Egerton 1992; Luo & Khursheed 2006). All these efforts so far have lacked detailed critical assessment of the measurement accuracy.

We have also attempted the aberration measurement and correction over several years, usually relying on a variety of *ad hoc* methods. In an attempt to collect and formalize some of these ideas, the current work describes our method for the aberration measurement and provides an assessment of the accuracy and limit of the aberration measurement. A simulation of the experiment based on wave mechanical formulae is introduced, which forms an essential part of the assessment. The assessment should be useful to guide further improvement of the measurement procedure, which can in turn lead to further progress in the energy resolution. The EELS aberration simulation may be also helpful for quantitative momentum-resolved EELS in the future. We will not address the correction of the measured aberrations in the present work.

**Theory**

**1. Energy resolution**

For EELS experiments, the spectrometer has two main functionalities: to disperse electrons of different energies; and to focus electrons of similar energy to the same channel. Ideally, electrons of the same energy (e.g., zero-energy loss electrons) are focused to a sharp line on the EELS detector. The width of a zero-loss peak (ZLP) defines the energy resolution, which depends on both the energy spread of the source image on the sample plane and aberrations of the post-sample optics (including projection lenses and the spectrometer). In monochromated STEMs, the energy spread of the source is greatly reduced by the monochromator, which contributes to a much better energy resolution of such instruments (Krivanek et al. 2009). To reach the optimal energy resolution in daily operations, automatic measurement and correction of the spectrometer aberrations become essential (Krivanek et al. 2019).

**2. Aberration**

Aberrations of optical systems are characterized by, in terms of geometric optics, deviation of rays from their ideal trajectories, which causes a focal point to spread and/or an image of an object to distort. From wave optical perspective, aberration can be formulated as the departure of wavefronts from their ideal shapes. The ideal wavefront of a STEM probe follows a paraboloid surface such that all the rays can be focused at one spot on the sample plane. The ideal wavefront for the zero-loss beam in EELS can be represented by a cylindrical surface (within the paraxial approximation), such that all the rays can be focused to a single line.

In STEM, chromatic aberration causes on-axial electron beams of different wavelengths to converge at different focal points along the optical axis and causes off-axial electron beams to converge at different points in the plane perpendicular to optic axis. In EELS, chromatic aberration is similar to the off-axial chromatic aberration in STEM except that a linear dispersion is created on purpose. The undesired second and higher order chromatic or mixed aberration in EELS can cause blurring peaks at high energy losses (MacLaren et al. 2018) even when the zero-loss peak is still in focus. For vibrational EELS, where the energy range of interest is small, the effect of chromatic and mixed aberrations can often be neglected, and geometric aberration is dominant.

Geometric aberration is a function of the distances from optic axis in the aperture plane (i.e., the momentum transfer) and causes monochromatic beams to intersect with the image plane at different points. This picture applies to both STEM and EELS. A pair of schematic diagrams in Figure 1 shows the effect of geometric aberration, which fails to bring all the monochromatic beams (of zero loss energy) to a focal line. The dispersive direction coincides with the x-direction along which geometric aberration causes spread of the beam. In the non-dispersive direction, y, an image of the diffraction spots is formed on the EELS detector plane. Without aberrations in the x-direction, the EELS detector plane is conjugate to the sample plane; in the y-direction, the EELS detector plane is conjugate to the diffraction plane (where the Ronchigram camera is located).

In addition to the geometric aberrations which limit the resolution, there is distortion, which does not affect resolution in the same way, but distorts the features of a finite object in an image. In optics textbooks, such as Born & Wolf, distortion is also listed as one type of geometric aberration, but in the present situation it is simpler to treat this separately. In momentum-resolved EELS, distortion happens primarily in the y-direction, and diffraction spots on the

EELS detector show disproportionate spacings, which might affect the accuracy of the measured dispersion relationships.

## 3. Expressions for geometric aberrations and distortion

First, we define the EELS aberration function, $\chi$, with a polynomial expansion:

$$\chi(\theta_x, \theta_y) = \sum_{n,m} \frac{E_{nm}}{n+1} \theta_x^{n+1} \theta_y^m \quad (1),$$

such that the x-gradient, which will be shown to be directly related to measurements, gives

$$\frac{\partial \chi(\theta_x, \theta_y)}{\partial \theta_x} = \sum_{n,m} E_{nm} \theta_x^n \theta_y^m \quad (2),$$

where $\theta_x$ and $\theta_y$ are the angles between the beam and the (curved) optic axis within the EELS aperture plane, $E_{nm}$ is the aberration coefficient, which is in length unit (so that the phase angle, $2\pi i \chi / \lambda$, would be dimensionless). Given a known dispersion setting, $E_{nm}$ can also be converted to energy units. While this conversion depends on the choice of dispersion settings, the aberration coefficients measured in length unit can be considered as a universal metric for different dispersion settings, since the EELS detector plane calibration can be converted as desired. Additionally, these aberrations in units of length unit can account for residual probe shifts during the alignment of the "ω-q" mode of momentum-resolved EELS. When evaluating the influence of geometric aberration on energy resolution, it might be convenient to work with units of energy.

The sum (n+m) characterizes the order of the aberration coefficients. For example, $E_{20}, E_{11}$ are second order aberrations and $E_{30}, E_{12}$ are third order ones.

Next, we define the polynomial expansion for distortion as:

$$\theta_y' = \sum_{n,m} D_{nm} \theta_x^n \theta_y^m \quad (3),$$

where $\theta_y'$ is the angle in the y-direction (i.e., the momentum direction on the EELS detector), $D_{nm}$ is the distortion coefficient, and $\theta_x$ and $\theta_y$ are the undistorted angles which are measured on the Ronchigram camera before the spectrometer. We ignore the magnification and set $D_{01} = 1$.

Alternatively to eq(3), the distortion can also be expressed by a function as follows:
$$\theta'_y = f(\theta_x, \theta_y) \quad (4).$$

If the distortion coefficients, $D_{nm}$, are included up to the second order, we can easily find an analytic expression for the inverse function
$$\theta_y = f^{-1}(\theta_x, \theta'_y) \quad (5)$$

through solving a quadratic equation. The purpose of having such an inverse function/transformation will be presented in Part 5 of the Experiment section.

## 4. Simulation of the aberration pattern

Here, we give the wave mechanical formulae for the aberration pattern, which is the intensity distribution of beams from different incident angles into the spectrometer.

$$\psi(x, k_y) = \int A(k_x, k_y) \exp\left[-i\frac{2\pi}{\lambda}\chi(k_x, k_y)\right] \exp[-2\pi i k_x x] \, dk_x \quad (6),$$

and

$$I(x, k_y) = |\psi(x, k_y)|^2 \quad (7),$$

where $k_x = \theta_x/\lambda$ and $k_y = \theta_y/\lambda$. $A(k_x, k_y)$ is the aperture function, which is 1 inside the aperture and 0 outside the aperture.

This form can be compared to a STEM probe at (x,y) where the intensity distribution is:

$$I(x, y) = |\psi(x, y)|^2$$
$$= \left| \iint A(k_x, k_y) \exp\left[-i\frac{2\pi}{\lambda}\chi(k_x, k_y)\right] \exp[-2\pi i(k_x x + k_y y)] dk_x dk_y \right|^2 \quad (8)$$

The analogy can be seen from Figure 2.

The results of simulated aberration patterns for a convergent beam of 30 mrad in semi-angle is shown in Figure 3. Each panel corresponds to the pure effect of a certain aberration coefficient.

To include distortion into the simulation, we perform an active transformation of the aberration function by substituting eq. (5) into eq. (6). This way of treating distortion in optics follows (Harres, Fuhrmann & Smith 1999).

The purpose for the simulation is to represent the aberration patterns with directly measurable quantities, i.e., the aberration and distortion coefficients. The simulation treats the aberration function as a 2D phase object and ignores the phase change with traveling along the optic axis.

## 5. The eikonal approximation-how the aberration coefficients are measured

One might think that measurement of aberration coefficients can be done by fitting simulated aberration patterns to the experimental ones. However, such a thought ignores the fact that such an optimization is almost computationally intractable, although machine-learning might present a promising solution. Similarly, the fit might not be unique for a single pattern. An alternative route must therefore be pursued.

By assuming that the electron wave propagating through the magnetic optics has a constant amplitude, we can derive the eikonal approximation from the solution to the Schrodinger equation, which gives

$$\frac{\partial \chi(\theta_x, \theta_y)}{\partial \theta_x} = -x \quad (9).$$

A detailed derivation of the dual form (where the spatial gradient of the phase equals the momentum) can be found in the quantum mechanics textbook by Steven Weinberg (2015). Historically, the eikonal equation is essential to Hamiltonian/geometric optics (Born & Wolf 2019). Early investigations of electrons optics by Rose also present the same formula from the perspective of Hamiltonian/geometric optics (Rose 1987). The left-hand side of eq. (9) is related to the aberration coefficients according to eq. (2), while the right-hand side is a measurable quantity from the EELS detector. This equation forms the guide for the experiment.

# Experiment

## 1. A brief description

We carried out the experiments in a monchromated-aberration-corrected Nion STEM (MACSTEM), at 60 kV. A semi-convergence angle of about 3 mrad was used for the momentum resolution, and the beam current was about 20 pA. To measure geometric aberrations, a wide angular range of incident beams was required. Thus, we set up an automatic beam tilt control while recording the beam on Ronchigram/EELS camera.

## 2. Purity of tilt

The 'purity' of beam tilt is important because tilt-induced beam shift on the sample plane is imaged to the EELS camera, which complicates the EELS aberration measurement. To ensure that the beam is tilted about a pivot point on the sample plane, the beam tilt needs to be aligned first. This alignment was done by manually tilting the beam (with descan switched on) and observing the annular dark field (ADF) STEM images of a few bright features in the sample (Figure 4). After this step, we were ready to collect data for the aberration measurements.

## 3. Automatic collection of the aperture images and ZLP images under the tilt series

We developed a simple python program and implemented it into Nionswift to interface with the Nion microscope. The program automates the beam tilts and serially records the beam tilts, $(\theta_x, \theta_y)$, on the Ronchigram camera. The same tilt series is then repeated while the ZLP image is recorded.

## 4. The procedure of extracting the aberration coefficients

Combining the Eikonal approximation, eq. (9) and the expansion, eq (2)

We have,

$$\sum_{n,m} E_{nm} \theta_x^n \theta_y^m = -x \quad (10).$$

The equation above can be written in a matrix form:

$$\begin{pmatrix} -x_1 \\ \vdots \\ -x_N \end{pmatrix} = \begin{pmatrix} (\theta_x^1 \theta_y^0)_1 & (\theta_x^0 \theta_y^1)_1 & \cdots & (\theta_x^n \theta_y^m)_1 \\ \vdots & & \ddots & \vdots \\ (\theta_x^1 \theta_y^0)_N & (\theta_x^0 \theta_y^1)_N & \cdots & (\theta_x^n \theta_y^m)_N \end{pmatrix} \begin{pmatrix} E_{10} \\ \vdots \\ E_{nm} \end{pmatrix} \quad (11)$$

Where $N$ is the total number of beam tilts/measurements and $x_N$ is the x-direction shift on the EELS camera for the $N^{th}$ beam tilt. We can also re-write the matrix formula in a shorthand notation as

$$-X\_{shift} = A\,E \quad (12),$$

where $A$ is the angular matrix, and $E$ is the column vector of aberration coefficients.

From this dataset, there are two obvious methods of extracting the aberration coefficients: 1) matrix inversion, 2) polynomial regression. For the matrix inversion, the following formula is used:

$$E = (A^T A)^{-1} A^T (-X\_{shift}) \quad (13)$$

For polynomial regression, the objective function

$$\sum_N [(\theta_x^1 \theta_y^0)_N E_{10} + \cdots (\theta_x^n \theta_y^m)_N E_{nm} - (-x_N)]^2 \quad (14)$$

is to be minimized. It should be noted that the expansion only contains linear terms of the aberration coefficients, and the angular terms are measurable quantities. Therefore, the polynomial regression here is a linear regression problem and admits a simple solution for the aberration coefficients.

## 5. The procedure of extracting the distortion coefficients

We extract the distortion coefficients up to the second-order terms using equation (3). In a matrix notation, equation (3) is rewritten as

$$\theta_y' = A\,D \quad (15)$$

where $\theta'_y$ is the distorted angle obtained from the EELS camera for the (undistorted) beam tilt $(\theta_x, \theta_y)$ from the Ronchigram camera; $A$ is the same angular matrix as in eq. (12); $D$ is the column vector of the distortion coefficients up to the second-order:

$$\boldsymbol{D}^T = (D_{01}, D_{02}, D_{10}, D_{11}, D_{20}) \quad (16).$$

Using these distortion coefficients, we obtain the expression for the distorted angles as

$$\theta'_y = f(\theta_x, \theta_y) \quad (17).$$

To include the distortion effect in the simulation, we cannot substitute $(\theta_x, \theta'_y)$ into eq. (6) directly because eq. (6) is the expression for the undistorted space. An inverse coordinate transformation must be performed first with the expression

$$\theta_y = f^{-1}(\theta_x, \theta'_y) \quad (18).$$

Then, together with the original value of $\theta_x$, we substitute $(\theta_x, f^{-1}(\theta_x, \theta'_y))$ into eq. (6) to model the wavefront in the distorted space. This method of incorporating the distortion effect into the wavefront simulation by using an inverse coordinate transformation follows the approach adopted in (Harres, Fuhrmann & Smith 1999).

### 6. Summary of the aberration measurement

An overview of the EELS aberration measurement is summarized in Figure 6. The aberration simulation in Figure 6 does not consider any distortion in the y-direction. The more complete simulation, including the distortion effect, will be presented in Figure 7(e).

### Results

We collected 3 datasets under different aberration conditions (Figs. 7a-c) and extracted their aberration and distortion coefficients according to the Experiment section. The first-order aberrations (E10 and E01) were tuned manually in Fig 7a. The first and second order aberrations were set large in the Fig 7b, and high order aberrations were set large in addition to the low-order aberrations in Fig 7c.

The three aberration patterns (Figs. 7a-c) share the same tilt series as shown in Figure 5 and 6, which spans from -24 mrad to +24 mrad in both directions. The simulation of the aberration patterns (Figs 7d-e) was carried out based on the extracted aberration and distortion coefficients. A good agreement between the simulation and experiment is observed for the second (Figs. 7 b&e) and third cases (Figs. 7c&f). The agreement in the first case, where the first-order aberrations were manually minimized, was inferior to the other two cases.

The distortion coefficients were measured and the scatter plots of ($\theta_x, \theta'_y$) summed over all the tilts is displayed in Figs. 7g-j. Clearly, the distortion depends on the aberration condition.

In fitting the aberration coefficients through the eikonal equation, eq.(9), we have used the weight-averaged beam shifts/phase gradients and angles. This practice inevitably causes errors. Here, we evaluate such errors by extracting aberration coefficients from simulated aberration patterns and compare them to the inputs. The results are summarized in Table 1. Two things can be noticed from the table. First, within the tilt range of (-24 mrad, +24 mrad) in both directions, accurate aberration measurement can be obtained up to (including) the third-order aberrations, whereas higher-order aberrations have large errors. Second, the Linear Regression algorithm shows better accuracy than the matrix inversion method (at least for the selected aberration conditions).

To explore the possibility of improving the accuracy of higher-order aberrations, we also attempted the measurement from simulated aberration patterns with more tilts and a bigger angular range. However, the extracted aberration coefficients of all orders show large departures from the inputs even though the total goodness of fit is 0.97.

**Discussion**

In the "ω-q" mode of momentum-resolved EELS, a slot aperture is used to select a column of diffraction spots (along the y-direction) and the diffraction pattern usually needs to be rotated to align with the slot. The rotation is carried out by adjusting projection lenses, which inevitably introduces extra aberrations and distortion at the final detector plane. During a session of momentum-resolved EELS, it is necessary to adjust the projection lenes upon moving to a new sample area with a different rotation, which requires the user to manually adjust the aberration coefficients. It seems clear that automatic aberration measurement and correction would greatly benefit this process.

The quality of aberration correction also relies on the accuracy of the measurement. We tested the accuracy of the aberration measurement under different aberration conditions by visual comparisons of the simulations and experiments. After finishing the purity of beam tilt, we can ignore the residual tilt-induced beam shift on the sample plane in our simulation, which still shows good agreement with the experiment. More delicate measurement may be achieved by including a measurement of the residual probe aberration from Figure 4. Results in Figs. 7a-f indicate that it seems better to carry out the aberration measurement under a defocused E10 (also known as Fx). This is conceivably because with a smaller defocus of E10, the small beam shift can be more affected by errors due to residual instabilities. Defocused E10 is often used in manual EELS tunning in STEM mode where a large convergence beam of illumination is used. Fig. 7g reveals that we should be alerted to the possibility of distortion even when the first-order aberration is minimized. The distortion could potentially be a serious issue in the experiment of momentum-resolved EELS with a 2D rocking beam or a shift of diffraction patterns by the projection lenses.

Table 1 reveals that the accuracy of measurement depends on the choice of fitting algorithm. The linear regression method, which includes statistical fluctuation into consideration, shows a better performance than the matrix inversion method. With 48 independent measurements of the beam shift (under 7x7 tilts), 9 aberration coefficients (up to the third-order) can be measured accurately. This over determinacy is required by linear regression, where a rule of thumb is that the number of data points should be at least 5 times the number of fitted parameters (Bertsekas & Tsitsiklis 2008).

Table 2 suggests that the fitted aberration coefficients may not be unique when a single polynomial form is used to describe aberration patterns with large beam tilts, and the values depend on the configuration of the beam tilts. This indeterminacy is not surprising, and the issue has been discussed in the literature on phase gradient measurement for radar (Herrmann 1981). Aberration aliasing always happen in practical measurements where higher-order aberrations become indistinguishable from lower-order aberrations when the phase-gradient is evaluated on a finite mesh. In addition to aberration aliasing, there is cross coupling between the lower and higher order coefficients due to the lack of orthogonality of the basis in the polynomial. Our choice of the polynomial basis follows the Nion convention for EELS aberration and is also similar to the CEOS convention. It should be noted that even though each individual aberration coefficient may have a large error, the summed polynomial, which gives the phase gradient, carries fairly accurate and meaningful information about the aberration condition. It may also be worth noting for future development in automatic correction that minimization of individual aberration coefficients in a single polynomial may not be very meaningful for fourth order and fifth order aberrations. First, second and third order terms can be measured and corrected within a small range and set fixed before carrying out fourth and fifth order measurement over a larger angular range, perhaps using a spline fitting scheme.

Our EELS aberration measurement is ultimately intended to be part of an automatic removal of the aberrations. Even though some automatic corrections of the spectrometer have been implemented by Gatan (Twesten *et al*. 2011) and CEOS (Kahl *et al.* 2019), more dedicated tunning may be needed for the Nion system with ultrahigh energy-resolution. The aberration correction in STEM can also guide our EELS tuning. In the automatic probe correction in STEM, removal of a certain aberration coefficient is controlled by a certain linear combination of multipole excitations (i.e., tuning currents in the multipole). Automatic EELS aberration correction may follow the same principle. However, the cross-coupling of aberration terms could present challenges to such a strategy. Similarly, using a linear combination of controls to minimize aberrations can only reach a certain degree of accuracy since this practice assumes that the phases at different optical component planes can be simply added and the nonlinear effect due to multiple scattering (Cowley & Moodie 1958) is ignored. To include such nonlinear effects into tuning, which may be essential for tuning high order aberrations, we may need simulations of the electron trajectories, for example based on the geometric optics formulation in the TRANSPORT code (Carey *et al.* 1995).

**Conclusion**

In summary, we have presented an automatic and accurate measurement of the spectrometer aberrations in a monochromated STEM. A simple simulation algorithm is proposed and has shown to be effective in providing a visually good agreement between the simulation and experiment. We have also assessed the accuracy of the aberration measurement and proposed strategies for improving the measurement of higher-order aberrations and for correction of the aberrations.

In principle the measurement process here could be extended to chromatic and off-axial terms. For example, the beam energy might be varied and a tilt series recorded at each energy. To speed up the process, perhaps fewer tilts could be used and more carefully selected, depending on the terms of interest.

Finally, we suggest that this wave optical formulation of the EELS aberrations may be useful for quantitative momentum-resolved EELS in the future. Just as consideration of the probe aberration function is useful for quantitative STEM, quantitative momentum-resolved EELS should include spectrometer aberrations and distortions.

**Code availability**

The codes for automatic data collection and measurement are available from

https://github.com/DrYGuo/EELS-aberration-measurements-and-simulations.

**Acknowledgement**

This work is supported by the U.S. Department of Energy (DOE), Office of Science, Basic Energy Sciences, Materials Sciences and Engineering Division and conducted at the Center for Nanophase Materials Sciences (CNMS), which is a US Department of Energy, Office of Science User Facility at Oak Ridge National Laboratory.

**Conflict of Interest**

The authors declare none.

**Figures**

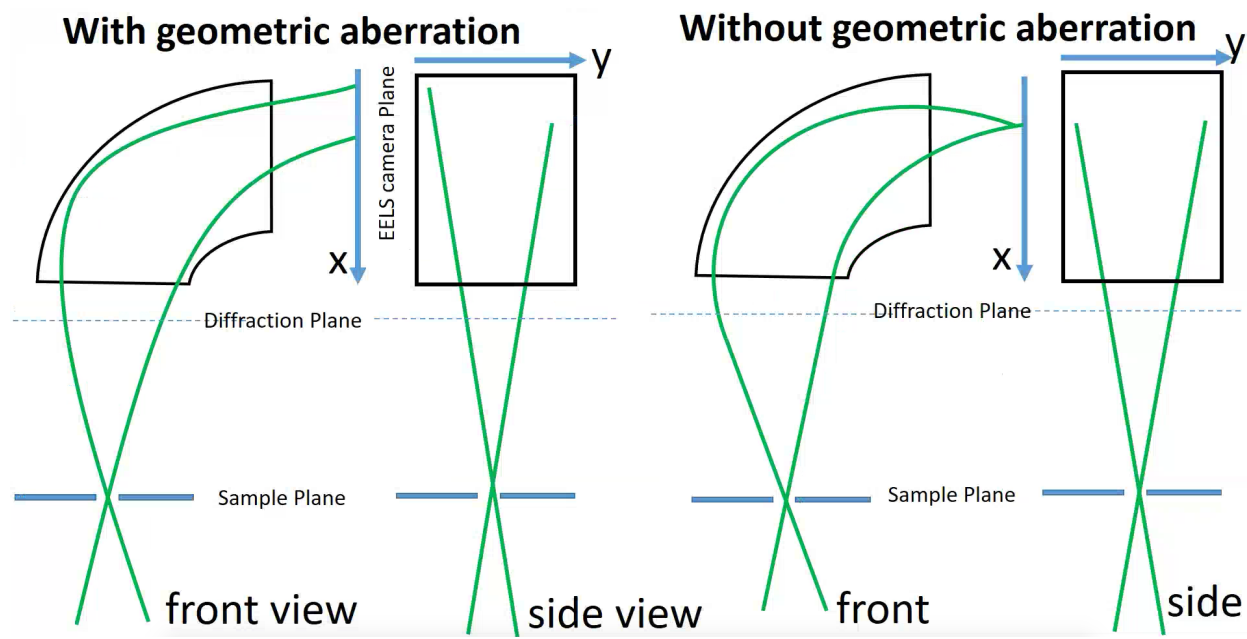

Figure 1. A schematic diagram that shows the effects of geometric aberration on electron trajectories in EELS. The effect of geometric aberration spreads the electron beam in the x direction (dispersion direction) even in the absence of a sample.

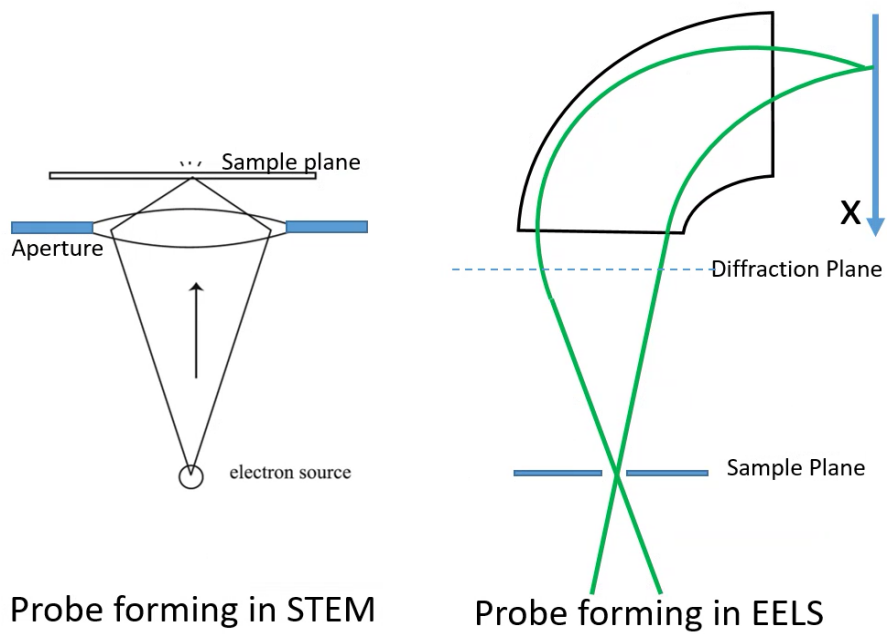

Figure 2. Analogy between the STEM probe formation and EELS probe formation.

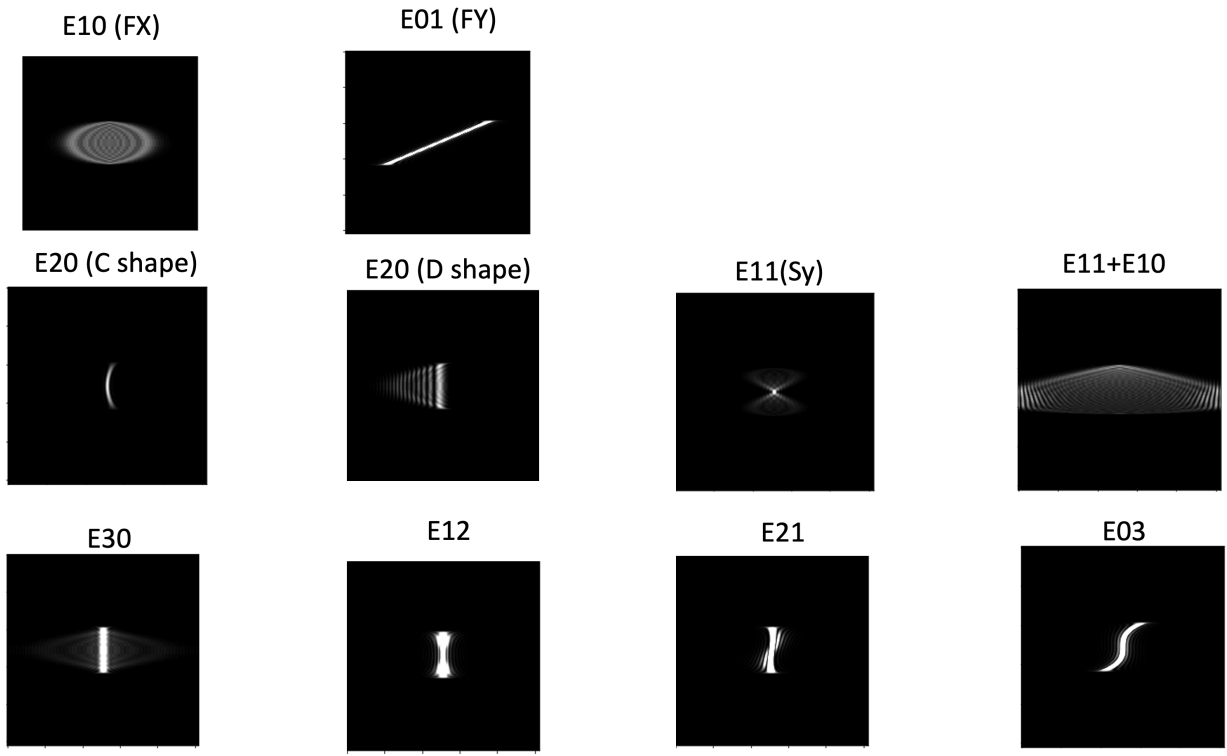

Figure 3. Simulated aberration patterns of all the geometric aberration coefficients up to the third order for a 30 mrad convergent beam at 60 kV.

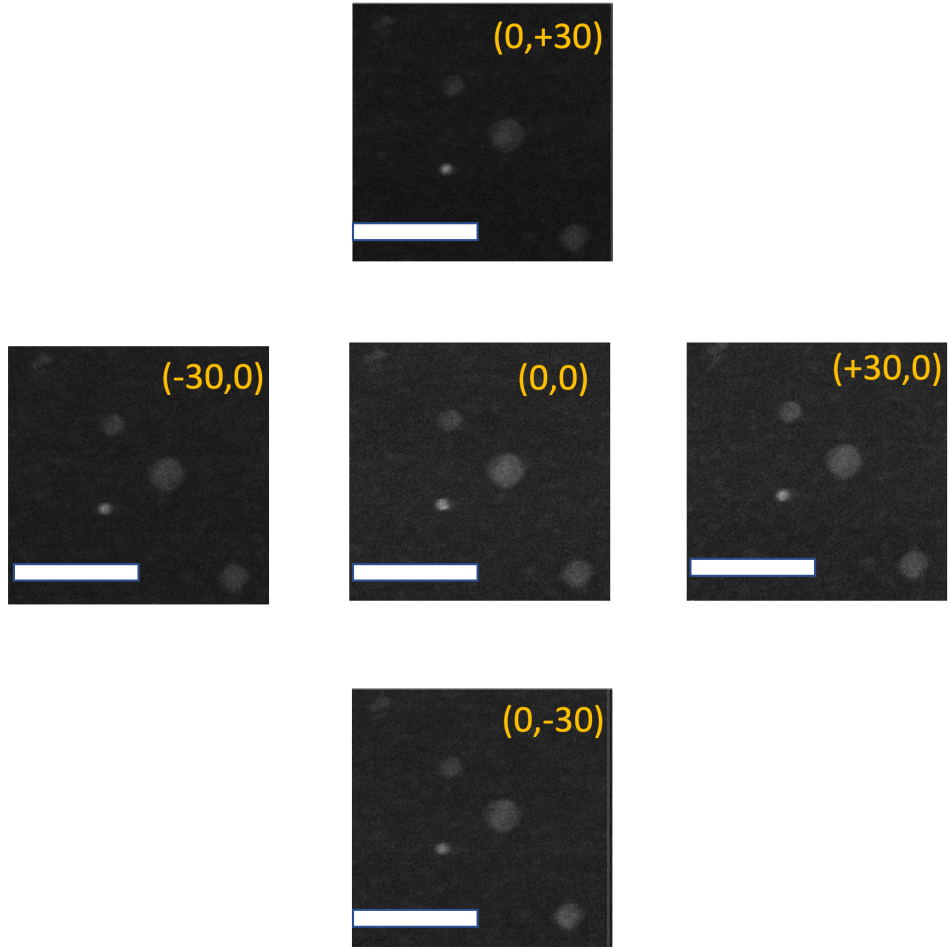

Figure 4. STEM images under beam tilts after adjusting purity of beam tilt. The scale bar represents 200 nm.

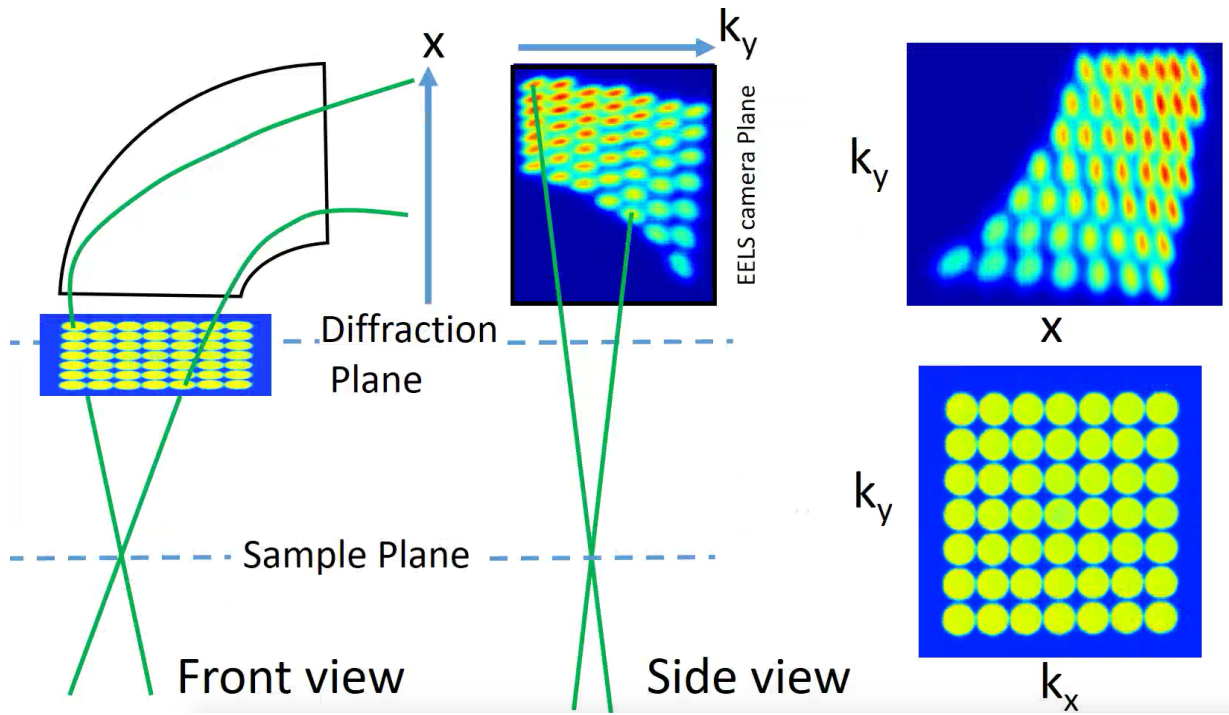

Figure 5. Illustration of the computer-controlled data collection. The computer code controls the beam tilt about a pivot point at the sample plane. The diffraction patterns (images of the condenser aperture) are recorded sequentially by the Ronchigram camera. The same tilt series is then repeated and the probe image on EELS camera is recorded. (Figure adapted from abstract (Guo and Lupini, 2022)).

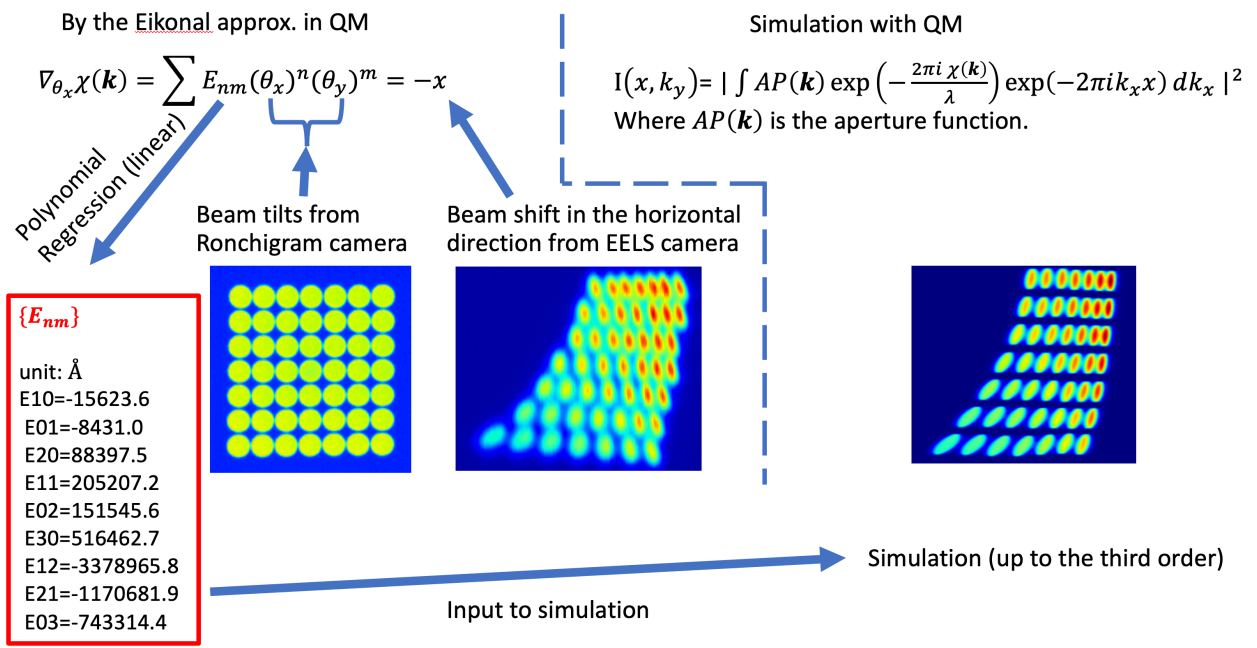

Figure 6. An overview of the EELS aberration measurement and verification via simulation. (Figure adapted from abstract (Guo and Lupini, 2022)).

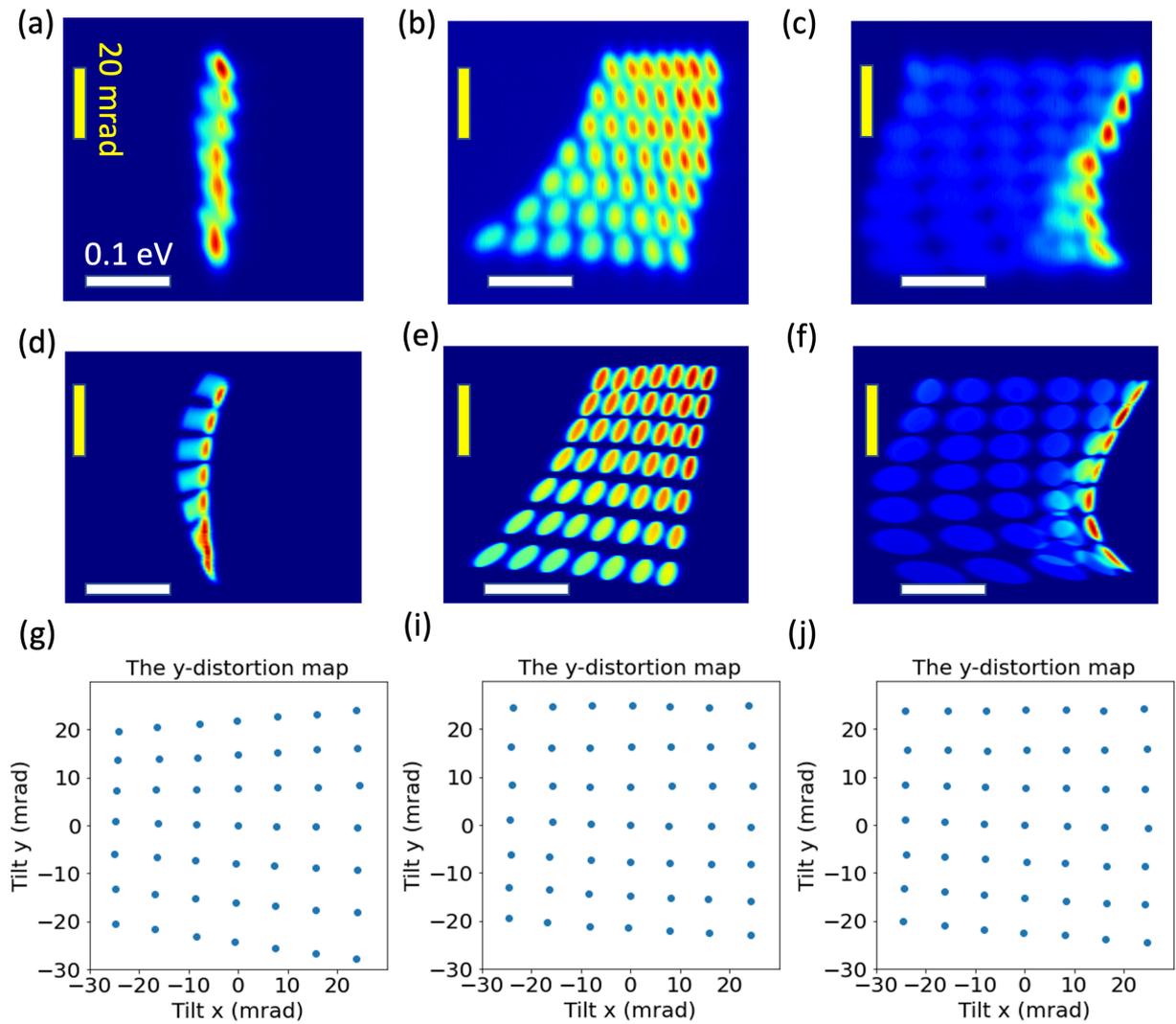

Figure 7. Comparison between experimental (a-c) and simulated (d-f) aberration patterns plus the (g-i) distortion map.

Table 1. Accuracy with two different algorithms for the aberration fitting (7x7 tilts, range +/- 24 mrad).

| Coefficients | Inputs (Angstrom) | Matrix operations | | LinearRegression in sklearn | |
|---|---|---|---|---|---|
| | | Measured values | Errors (%) | Measured values | Errors (%) |
| E10 | -15617.7 | -15598.7 | -0.12166 | -15596.9 | -0.13318 |
| E01 | -8462.6 | -8469.3 | 0.079172 | -8460.5 | -0.02482 |
| E20 | 48975.4 | 59689.8 | 21.87711 | 49189.0 | 0.436137 |
| E11 | 183778.0 | 184140.1 | 0.197031 | 183851.0 | 0.039722 |
| E02 | 69443.8 | 81129.1 | 16.82699 | 69620.9 | 0.255026 |
| E30 | 463585.4 | 486851.2 | 5.018665 | 482260.8 | 4.02847 |
| E12 | -3221234.6 | -3183973.5 | -1.15673 | -3180764.6 | -1.25635 |
| E21 | -1221098.3 | -1166466.9 | -4.47396 | -1180412.7 | -3.33189 |
| E03 | -577915.4 | -540653.1 | -6.44771 | -561938.6 | -2.76456 |
| E40 | 44436756.8 | -11575925.9 | -126.05 | -474250.9 | -101.067 |
| E13 | 89672210.6 | -1578776.8 | -101.761 | -1131933.5 | -101.262 |
| E22 | -28323867.6 | -10431437.5 | -63.1709 | -1618846.7 | -94.2845 |
| E31 | -33327076.7 | -1638316.3 | -95.0841 | -1270955.8 | -96.1864 |
| E04 | 119603183.5 | -13843843.7 | -111.575 | -473556.2 | -100.396 |

Table 2. Indeterminacy of the exact aberration coefficients (13x13 tilts, range +/-40 mrad).

| Coefficients | Inputs (Angstrom) | Matrix operations | | LinearRegression in sklearn | |
|---|---|---|---|---|---|
| | | Measured values | Errors (%) | Measured values | Errors (%) |
| E10 | -15617.7 | -17142.6 | 9.763922 | -17142.6 | 9.763922 |
| E01 | -8462.6 | -9949.4 | 17.56907 | -9949.4 | 17.56907 |
| E20 | 48975.4 | -118928.7 | -342.834 | -35115.9 | -171.701 |
| E11 | 183778.0 | 276155.2 | 50.26565 | 275884.2 | 50.11819 |
| E02 | 69443.8 | 28574.9 | -58.8518 | 112387.7 | 61.83979 |
| E30 | 463585.4 | 4976721.5 | 973.5285 | 4976721.5 | 973.5285 |
| E12 | -3221234.6 | -5459017.6 | 69.46973 | -5467148.0 | 69.72213 |
| E21 | -1221098.3 | 782137.7 | -164.052 | 790268.2 | -164.718 |
| E03 | -577915.4 | -442390.1 | -23.4507 | -442390.1 | -23.4507 |
| E40 | 44436756.8 | 202231430.3 | 355.0994 | 169221833.4 | 280.815 |
| E13 | 89672210.6 | -133621508.5 | -249.011 | -133621508.5 | -249.011 |
| E22 | -28323867.6 | -100606828.6 | 255.2016 | -129203812.6 | 356.1659 |
| E31 | -33327076.7 | 92719142.5 | -378.21 | 92719142.5 | -378.21 |
| E04 | 119603183.5 | 52640223.7 | -55.9876 | 19630626.8 | -83.5869 |